\documentclass[12pt,a4paper]{article}

\usepackage[english]{babel}
\usepackage{epsfig}
\usepackage{amsmath}
\usepackage{amssymb}
\usepackage{amsbsy}
\usepackage{amsfonts}
\usepackage{array}
\usepackage{verbatim}
\usepackage{graphicx}
\usepackage{subfigure}
\usepackage{tabularx}
\usepackage{bm}
\usepackage{booktabs}
\usepackage[font=small]{caption}
\usepackage{stmaryrd}

\newcommand{\mG}{\mathcal{G}}

\newcommand{\mR}{\mathcal{R}}
\newcommand{\mS}{\mathcal{S}}
\newcommand{\mT}{\mathcal{T}}

\newcommand{\mV}{\mathcal{V}}

\newcommand{\Lij}{\mathcal{L}_{ij}}
\newcommand{\Lkk}{\mathcal{L}_{kk}}
\newcommand{\Li}{\mathcal{L}_{i}}

\newcommand{\Sij}{\mathcal{S}_{ij}}
\newcommand{\Skk}{\mathcal{S}_{kk}}

\newcommand{\bu}{\mathbf{u}}

\newcommand{\bF}{\mathbf{F}}

\newcommand{\bmG}{\boldsymbol{\mG}}

\newcommand{\bS}{\mathbf{S}}

\newcommand{\bU}{\mathbf{U}}

\newcommand{\bq}{\mathbf{q}}

\newcommand{\bbar}{\overline}

\newcommand{\frho}{\bbar{\rho}}

\newcommand{\fp}{\bbar{p}}

\newcommand{\wt}{\widetilde}

\newcommand{\fu}{\wt{u}}
\newcommand{\fh}{\wt{h}}
\newcommand{\fe}{\wt{e}}

\newcommand{\fT}{\wt{T}}

\newcommand{\fS}{\wt{\mathcal{S}}}
\newcommand{\fSij}{\wt{\mS}_{ij}}

\newcommand{\fSkk}{\wt{\mS}_{kk}}

\newcommand{\fsigmaij}{\wt{\sigma}_{ij}}

\newcommand{\wh}{\widehat}

\newcommand{\de}{\partial}
\newcommand{\Div}{\nabla\cdot}

\newcommand{\sigmaij}{\sigma_{ij}}
\newcommand{\taukk}{\tau_{kk}}
\newcommand{\tauij}{\tau_{ij}}
\newcommand{\ei}{e_{\rm i}}

\newcommand{\hfu}{\breve{\fu}}
\newcommand{\hfT}{\breve{\fT}}
\newcommand{\hfrho}{\widehat{\frho}}
\newcommand{\hfS}{\breve{\fS}}

\newcommand{\bn}{\mathbf{n}}
\newcommand{\bx}{\mathbf{x}}
\newcommand{\br}{\mathbf{r}}

\newcommand{\hdelta}{\widehat{\Delta}}

\newcommand{\deltaij}{\delta_{ij}}
\newcommand{\nusgs}{\nu^{{\rm sgs}}}

\newcommand{\RE}{Re}
\newcommand{\MA}{M\hspace{-1pt}a}
\newcommand{\PR}{Pr}
\newcommand{\dm}{{\rm d}}

\begin{document}

\title{Large Eddy Simulation of gravity currents with a high order DG method}
\author{Caterina Bassi$^{(1)}$\\
 Antonella Abb\`a $^{(2)}$, Luca Bonaventura $^{(1)}$, Lorenzo Valdettaro $^{(1)}$}
\maketitle

\begin{center}
{\small
$^{(1)}$ MOX -- Modelling and Scientific Computing, \\
Dipartimento di Matematica, Politecnico di Milano \\
Via Bonardi 9, 20133 Milano, Italy\\
{\tt caterina.bassi@polimi.it, luca.bonaventura@polimi.it, lorenzo.valdettaro@polimi.it }
}
{\small
$^{(2)}$ Dipartimento di Scienze e Tecnologia Aerospaziali, Politecnico di Milano \\
Via La Masa 34, 20156 Milano, Italy\\
{\tt antonella.abba@polimi.it}
}
\end{center}

\date{}


\vspace*{0.5cm}


\vspace*{0.5cm}

\pagebreak


\abstract{This work deals with Direct Numerical Simulations (DNS) and Large Eddy Simulations (LES) of turbulent gravity currents, performed by means of a Discontinuous Galerkin (DG) Finite Element method.
In particular, a DG-LES approach in which the filter operation is built in the numerical discretization has been employed, similarly to VMS approaches.
Numerical simulations of non-Boussinesq lock-exchange benchmark problems show that, in the DNS case, the proposed method allows to correctly reproduce relevant features of variable density flows with gravity. Moreover LES results highlight the excessively high dissipation of the Smagorinsky model with respect to the Germano dynamic procedure, providing a first indication of the superiority of dynamic models in the context of gravity currents.}

\pagebreak
 
\section{Introduction}
\label{intro} \indent

Gravity currents are frequently encountered in geophysical flows,
when a heavier fluid propagates into a lighter one in a predominantly horizontal direction because of the difference in hydrostatic pressure at the boundary between the two fluids.
 In atmospheric gravity currents, such as thunderstorm outflows, density difference is typically caused by the temperature difference between the cold front and the warmer surrounding air. In oceanic flows, density difference is caused by    salinity and temperature gradients, while in pyroclastic flows the density difference is due to the presence of suspended particles in the flow. The evolution of gravity currents is also very important  
  for some engineering applications, such as the accidental leakage of industrial gases, see e.g. \cite{houf:2008}.
  
  In gravity currents, the density difference between the lighter and heavier fluid
  can range from very small to very large. In case of small density differences, density variations
  in the momentum equation can be neglected in the inertia term, but retained in the buoyancy term. This approximation is called Boussinesq approximation, see e.g. \cite{gill:1982}, and is sufficiently accurate for
  density differences up to a   few percent. In most experimental and computational studies of gravity currents reported in the literature, the
  Boussinesq approximation has been employed. However, in several of the above listed phenomena
   non-Boussinesq effects become important. 
  
  In the present work, we validate   the Large Eddy Simulation model recently proposed in \cite{abba:2015}
  for application to the study of turbulent gravity currents. This model
 is based on a high order, Discontinuous Galerkin finite element discretization. Such a  numerical
 framework allows to generalize the concept of LES filter as a projection onto the polynomial space related to the discretization, thus making it possible to  apply it to arbitrary unstructured meshes.
Furthermore, we model gravity currents via the Navier-Stokes equations for a compressible fluid in the limit of very low Mach number, thus  allowing for the simulation of both Boussinesq and non-Boussinesq gravity currents.  
We remark that, even if some   LES studies of Boussinesq gravity currents have  already been presented in the literature,  see e.g. \cite{ozgokmen:2011}, \cite{ozgokmen:2009}, \cite{ozgokmen:2006},  to the best of our knowledge there are almost no published results  in the non-Boussinesq case. 
We first provide an assessment of the ability of our DG-LES model to reproduce the incompressible results obtained in \cite{birman:2005}. The benchmark which has been taken into consideration is the canonical lock-exchange test case. This benchmark appears to be particularly interesting because it has been widely investigated both experimentally and numerically and, moreover, it provides a quite complex flow evolution (with the presence of shear driven mixing and internal waves) while being specified by  simple and unambiguous forcing, initial and boundary conditions, see e.g. the discussion in \cite{ozgokmen:2006}. We consider the lock-exchange benchmark in the non-Boussinesq regime, carrying out first simulations where all the turbulent scales of motion are correctly captured by the computational grid (Direct Numerical Simulations). We then present results of simulations at higher Reynolds number (Large Eddy Simulations) obtained with the classic Smagorinsky model and with the Germano dynamic model. The numerical results reported in section \ref{lock}  show that the method proposed in \cite{abba:2015} is able to reproduce correctly all the relevant features of this important benchmark.

The paper is organized as follows. In section \ref{model}, the general model problem is introduced. Section \ref{turbulence_model} is devoted to the description of the turbulence models that have been applied. 
The space and time discretizations employed and their link with the proposed turbulence modeling approaches are presented in section \ref{dgmeth}. The results of the numerical simulations are presented in section \ref{lock}, while some conclusions and perspectives for future work are discussed in section \ref{conclu}.

\section{The model problem}
\label{model}
We consider the compressible Navier--Stokes equations,
which can be written in dimensional
form (denoted by the superscript ``*''), employing the Einstein notation, as:
\begin{subequations}
\label{eq:nscompr-dim}
\begin{align}
&\de_{t}^* \rho^* + \de_{x_j}^* (\rho^* u^*_j) = 0, \\
&\de_{t}^* (\rho^* u^*_i) + \de_{x_j}^* (\rho^* u^*_i
 u^*_j) + \de_{x_i}^* p^* - \de_{x_j}^* \sigmaij^*  \nonumber \\
&\hspace{3cm} = \rho^* f^*_i, \\
&\de_{t}^* (\rho^* e^*) + \de_{x^*_j} (\rho^* h^* u^*_j)
- \de_{x_j}^* (u^*_i \sigmaij^*) + \de_{x_j}^* q^*_j  \nonumber \\
&\hspace{3cm} = \rho^* f_j^* u_j^*,
\end{align}
\end{subequations}
where $\rho^*$, $\bu^*$ and $e^*$ denote density, velocity and
specific total energy, respectively, $p^*$ is the pressure,
$\mathbf{f}^*$ is a prescribed forcing (in the present investigation the gravity forcing is considered and we have $\mathbf{f}^*=(0,0,-g^*)$), $h^*$ is the specific
enthalpy, defined by $\rho^* h^*=\rho^* e^*+p^*$, and
$\sigma^*$ and $\bq^*$ are the diffusive momentum and heat fluxes.
Equation~(\ref{eq:nscompr-dim}) must be complemented with the state
equation
\begin{equation}
p^* = \rho^* R^* T^*,
\label{eq:state-eq-dim}
\end{equation}
where $T^*$ is the temperature and $R^*$ is the ideal gas constant.
The temperature can then be expressed in terms of the prognostic
variables introducing the specific internal energy $\ei^*$, so that
\begin{equation}
e^* = \ei^* + \frac{1}{2}u^*_ku^*_k ,
\qquad T^* = \frac{\ei^*}{c_v^*},
\label{eq:ei-dim}
\end{equation}
where $c_v^*$ is the specific heat at constant volume. Finally, the
model is closed with the constitutive equations for the diffusive
fluxes:
\begin{equation}
\sigmaij^* = \mu^* \Sij^{d,*}, \qquad
q^*_i = -\frac{\mu^* c_p^*}{\PR} \de_{x_i}^* T^*,
\label{eq:constitutive-dim}
\end{equation}
where $\Sij^* = \de_{x_j}^* u_i^* + \de_{x_i}^*
u^*_j$ and $\Sij^{d,*} = \Sij^* -
\dfrac{1}{3}\Skk^*\delta_{ij}$,  $c_p^*=R^*+c_v^*$ is the specific heat at constant
pressure, $\PR$ denotes the Prandtl number,
and the dynamic viscosity $\mu^*$ is assumed to depend only on
temperature $T^*$ according to the power law
\begin{equation}\label{eqn:mu_def-dim}
\mu^*(T^*) = \mu^*_0 \left(\dfrac{T^*}{T^*_0}\right)^{\alpha}
\end{equation}
following Sutherland's hypothesis (see e.g. \cite{schlichting:1979}). 
The dimensionless form of the problem is obtained assuming four reference
quantities: $\rho_r$, $L_r$, $V_r$ and $T_r$. All the other reference quantities are derived from these fundamental ones by means of dimensional considerations (see Table \ref{table_one}).
\begin{table}\footnotesize
\centering
\begin{tabular}{ccc}
\toprule
Physical  & Reference & Fundamental\\
quantities & physical & physical \\
           & quantities&quantities \\
\hline
Length & $L_r$  & -    \\ 
Density & $\rho_r$ & -  \\
Velocity & $V_r$ & -\\
Temperature & $T_r$ & - \\
Pressure & $p_r$ & $\rho_r V_r^2$ \\
Gas constant & $R_r$ & $V_r^2/T_r$\\
Time & $t_r$ & $L_r/V_r$ \\
Internal energy & $e_r$ & $V_r^2$ \\
Dynamic viscosity & $\mu_r$ & $\rho_r L_r V_r$ \\
Thermal conductivity & $\lambda_r$ & $\rho_r V_r^3 L_r/T_r$ \\
Specific heat (const. vol.) & $c_{v,r}$ & $V_r^2/T_r$ \\
Specific heat (const. press.) & $c_{p,r}$ & $V_r^2/T_r$ \\
\bottomrule
\end{tabular}
\caption{Reference physical quantities} 
\label{table_one}
\end{table} 
Because of the fact that the number of the reference quantities is the same as the number of fundamental dimensions, the form of the dimensional and non-dimensional equations is the same:
\begin{subequations}
\label{eq:nscompr}
\begin{align}
&\de_t \rho + \de_j (\rho u_j) = 0, \\
&\de_t (\rho u_i) + \de_j (\rho u_i u_j) + 
\de_i p - \de_j \sigmaij =
\rho f_i,\label{eq:nscompr-momentum} \\
&\de_t (\rho e) + \de_j (\rho h u_j)
-  \de_j (u_i \sigmaij) \nonumber \\
&\hspace{3cm}+ \de_j q_j = \rho f_j u_j,   \label{eq:nscompr-energy},
\end{align}
\end{subequations}
where $\mathbf{f}=(0,0,-1/Fr^2)$. $Fr$ is the Froude number, which is related to the non-dimensional gravity acceleration by the following equation:
\begin{equation}
g = \frac{g^*}{g_r} = \frac{g^* L_r}{V_r^2} = \frac{1}{Fr^2}.
\end{equation}
Some other important quantities and equations in non-dimensional form are derived in the following. 
The non-dimensional form of the gas constant is given by:
\begin{equation}
R = \frac{R^*}{R_r} = \frac{R^* T_r}{V_r^2} = \frac{1}{\gamma \MA^2},
\label{adim_R}
\end{equation}
where $\MA$ is the Mach number and $\gamma$ the ratio between specific heats. 
The non-dimensional specific heat at constant volume is: 
\begin{equation}
c_v = \frac{c_v^*}{c_{v,r}} = \frac{c_v^* T_r}{V_r^2} = \frac{R^* T_r}{(\gamma -1) V_r^2} = \frac{1}{\gamma (\gamma-1) \MA^2}.
\label{adim_cv}
\end{equation}
Using equations (\ref{adim_R}) and (\ref{adim_cv}) we obtain for the specific heat at constant pressure:
\begin{equation}
c_p = R+c_v = \frac{1}{\gamma \MA^2} + \frac{1}{\gamma (\gamma-1) \MA^2} = \frac{1}{(\gamma -1) \MA^2}.
\end{equation}
The non-dimensional form of the state equation is obtained starting from (\ref{eq:state-eq-dim}).
Expressing the dimensional quantities as a function of the non-dimensional and reference ones we have:
\begin{equation}
\rho_r V_r^2 p = \rho_r \rho \frac{V_r^2}{T_r} R T_r T ,
\end{equation}
dividing by $\rho_r V_r^2$ and using equation (\ref{adim_R}) we obtain:
\begin{equation}
p= \rho R T = \frac{\rho T}{\gamma \MA^2}.
\label{ideal_gas_law}
\end{equation}
Starting from the dimensional equation (\ref{eq:ei-dim}), the non-dimensional internal energy is obtained as:
\begin{equation}
e_i = \frac{T}{(\gamma -1)\gamma \MA^2},
\label{ideal_gas_law_energy}
\end{equation}
while the total energy is:
\begin{equation}
e = \ei + \frac{1}{2} u_ku_k.
\label{total_energy_adim}
\end{equation}
The dynamic viscosity $\mu$ is given by:
\begin{equation}
\mu = \frac{\mu^*}{\mu_r} = \frac{\mu_0^*}{\rho_r L_r V_r} \biggl(\frac{T_r T}{T_0^*}\biggr)^\alpha = \frac{1}{Re} T^\alpha,
\label{dyn_visc}
\end{equation}
where the last equality holds because we set $T_0^\dm=T_r$. 
In \cite{birman:2005} the non-dimensional dynamic viscosity is:
\begin{equation}
\mu = \frac{\rho}{\RE}
\label{dyn_visc_birman}
\end{equation}
Since in our simulations the pressure field is substantially constant during the computation, this dependency upon density can be reproduced by retaining the Sutherland law (\ref{dyn_visc}) and by setting the parameter $\alpha = -1$. Alternatively equation (\ref{dyn_visc_birman}) can be directy employed. 
The thermal conductivity $\lambda$ is: 
\begin{equation}
\lambda = \frac{\mu c_p}{\PR}.
\end{equation}
The non-dimensional constitutive equations for the diffusive fluxes are: 
\begin{equation}
\sigmaij = \mu \Sij^d, \qquad
q_i = -\lambda\de_i T,
\label{eq:constitutive}
\end{equation}
with $\Sij = \de_j u_i + \de_i u_j$ and $\Sij^d = \Sij - \dfrac{1}{3}\Skk\delta_{ij}$.



\section{The LES model}
\label{turbulence_model}

We will present here a compact description of the LES models employed in this paper.
We refer to \cite{sagaut:2006} for a general introduction to LES modelling and to
\cite{abba:2015} for a more complete description of these and other LES models.
As it is well known, the key ingredient of a LES model is the filtering operator. 
In the approach proposed in \cite{abba:2015}, the filter operator is embedded in the
spatial DG discretization. The details of this realization of the filter operator will be 
given in section \ref{dgmeth}.
 Here, we only anticipate that the filtering operator is denoted by $\bbar{\cdot}$ and that it is associated with the spatial scale $\Delta$. The spatial scale depends on the local element size and is as a consequence a piecewise constant function in space. 
 
As customary in LES of compressible flows, see e.g. \cite{garnier:2009}, we introduce also the Favre filtering operator $\wt{\cdot}$ which is defined implicitly by the Favre decomposition. Given a generic function $f,$ the Favre decomposition is defined as:
\begin{equation}
\bbar{\rho f} = \frho \wt{f}
\end{equation} 
This decomposition is introduced for velocity, total energy, internal energy, enthalpy and temperature,
yielding the equations
\begin{subequations}
\label{ffilter_collective}
\begin{align}
\bbar{\rho u_i} & = \frho \wt{u}_i, \\
\bbar{\rho e} & = \frho \wt{e} = \frho \wt{e_i} +\frac{1}{2}\biggl(\bbar{\rho} \wt{u}_k \wt{u}_k + \tau_{kk}  \biggr), \label{ffilter_e}\\
\bbar{\rho e_i} & = \frho \wt{e_i} = \frac{1}{\gamma(\gamma-1)\MA^2}\bbar{\rho}\wt{T}, \label{ffilter_ei} \\
\bbar{\rho h} & = \frho \wt{h} = \frho \wt{e} + \bbar{p}, \\
\bbar{\rho T} & = \frho \wt{T} = \gamma \MA^2 \bbar{p}, \label{ffilter_T}
\end{align}
\end{subequations}
where the last equality in (\ref{ffilter_e}) holds because of equation (\ref{total_energy_adim}) while the last equalities in equations (\ref{ffilter_ei}) and (\ref{ffilter_T}) follow from the state equation written in the usual form (eq. (\ref{ideal_gas_law})) and in the internal energy form (eq. (\ref{ideal_gas_law_energy})) respectively. Notice that $\tau_{kk}$ in equation (\ref{ffilter_e}) is the trace of the subgrid stress tensor which is defined as:
\begin{equation}
  \tauij = \bbar{\rho u_i u_j} - \frho\fu_i\fu_j.
\label{eqn:tauij_sgs}
\end{equation}
We introduce also the filtered counterpart of the diffusive fluxes:
\begin{equation}
\fsigmaij = \mu \fSij^d, \qquad
\wt{q}_i = -\lambda \de_i \fT,
\label{eq:constitutive-Favre}
\end{equation} 
with $\fSij = \de_j \fu_i + \de_i \fu_j$ and $\fSij^d = \fSij -
\dfrac{1}{3}\fSkk\delta_{ij}$. 
Given these definitions, the filter operator is applied to the non-dimensional form of the Navier-Stokes equations (\ref{eq:nscompr}), thus obtaining:
\begin{subequations}
\label{eq:nscompr-filtered-intermediate}
\begin{align}
&\de_t \frho + \de_j (\frho \fu_j) = 0 \\
&\de_t \left( \frho \fu_i \right) + \de_j \left(\frho \fu_i \fu_j\right) 
+ \de_i \fp - \de_j \fsigmaij 
\nonumber \\
& \qquad \qquad
= - \de_j \tauij - \de_j \epsilon^{{\rm sgs}}_{ij} + \frho f_i \\
& \de_t \left(\frho\fe\right) + \de_j \left(\frho\fh \fu_j\right) 
- \de_j \left(\fu_i \fsigmaij \right)
+ \de_j \wt{q}_j \nonumber \\
& \qquad \qquad =
- \de_j \left(\rho h u_j\right)^{{\rm sgs}}
+ \de_j \phi^{{\rm sgs}}_j \nonumber \\
&\quad\quad\quad - \de_j \theta^{{\rm sgs}}_j + \frho f_j \fu_j,
\end{align}
\end{subequations}
where
\begin{equation}
\begin{array}{ll}
\epsilon^{{\rm sgs}}_{ij} = \bbar{\sigma}_{ij} - \fsigmaij, &
 \left(\rho h u_i\right)^{{\rm sgs}} = \bbar{\rho hu_i} -
 \frho\fh\fu_i,
\\[2mm]
 \phi^{{\rm sgs}}_j = \bbar{u_i \sigmaij} - \fu_i\fsigmaij, &
 \theta^{{\rm sgs}}_i = \bbar{q}_i - \wt{q}_i.
\end{array}
\label{eqn:epsij_sgs}
\end{equation}
Following  \cite{abba:2015}, we also make the  the assumptions $\bbar{\sigma}_{ij} \approx \fsigmaij$ and $\bbar{q}_i \approx \wt{q}_i.$ Furthermore, the term $\de_j \phi^{{\rm sgs}}_j$ is considered to be negligible, as
well as $\epsilon^{{\rm sgs}}_{ij}$ and $\theta^{{\rm sgs}}_j$. 
The treatment of the subgrid enthalpy flux $ \left(\rho h u_i\right)^{{\rm sgs}}$ is explained in the following. We notice that, by employing equations (\ref{total_energy_adim}), (\ref{ideal_gas_law_energy}) and (\ref{ffilter_T}) we obtain:
\begin{equation}
\frho\wt{h}  = \frac{1}{(\gamma -1)\MA^2}\frho \wt{T}
+ \frac{1}{2}\left( \frho\fu_k\fu_k + \taukk \right).
\end{equation}
Introducing the subgrid heat and turbulent diffusion fluxes:
\begin{subequations}
\begin{align}
Q_i^{{\rm sgs}} & = \bbar{\rho u_i T} - \frho\fu_i\fT =
 \frho \left( \wt{u_i T} - \fu_i\fT \right), \label{eqn:Qj_sgs} \\
J_i^{{\rm sgs}} & = \bbar{\rho u_iu_ku_k} - \frho \fu_i\fu_k\fu_k =
\frho \wt{u_i u_k u_k} - \frho\fu_i\fu_k\fu_k,
 \label{eqn:Jj_sgs}
\end{align}
\end{subequations}
we have:
\begin{equation}
\left(\rho h u_i\right)^{{\rm sgs}} = \frac{1}{(\gamma-1)\MA^2} Q_i^{{\rm sgs}}
+ \frac{1}{2} \left( J_i^{{\rm sgs}} - \taukk\fu_i \right).
\label{eq:enthalpy-sgs}
\end{equation}
Introducing the generalized central moments $\tau(u_i,u_j,u_k)$ \cite{germano:1992} with:
\begin{subequations}
 \begin{align}
\tau(u_i,u_j,u_k) = & \frho\wt{u_i u_j u_k} - \fu_i\tau_{jk} 
  - \fu_j\tau_{ik} - \fu_k\tauij \nonumber \\
& - \frho\fu_i\fu_j\fu_k,
\end{align}
\end{subequations}
$J_i^{sgs}$ in eq. (\ref{eqn:Jj_sgs}) can be rewritten as:
\begin{equation}\label{eqn:Jj_sgs2}
 J_i^{{\rm sgs}} = \tau(u_i,u_k,u_k) + 2\fu_k\tau_{ik} + \fu_i\taukk.
\end{equation}
The filtered equations (\ref{eq:nscompr-filtered-intermediate}) become:
\begin{subequations}
\label{filteq}
\begin{align}
&\de_t \frho + \de_j (\frho \fu_j) = 0 \\
&\de_t \left( \frho \fu_i \right) + \de_j \left(\frho \fu_i \fu_j\right) 
+ \de_i \fp - \de_j \fsigmaij \nonumber \\
& \qquad \qquad = - \de_j \tauij + \frho f_i \label{filteq-momentum} \\
& \de_t \left(\frho\fe\right) + \de_j \left(\frho\fh \fu_j\right) 
- \de_j \left(\fu_i \fsigmaij \right)
+ \de_j \wt{q}_j  \nonumber  \\
& \qquad \qquad =
- \frac{1}{(\gamma -1)\MA^2}\de_j Q_j^{{\rm sgs}}
- \frac{1}{2}\de_j \left( J_j^{{\rm sgs}} - \taukk\fu_j \right)  \label{filteq-energy} \\
&\quad\quad\quad + \frho f_j \fu_j.   \nonumber 
\end{align}
\end{subequations}

The term $\tau_{ij}$ in  the momentum   equation (\ref{filteq-momentum}) and $Q_j^{{\rm sgs}}$ and $J_j^{{\rm sgs}}$ in energy conservation equation (\ref{filteq-energy}) are subgrid terms and need modeling. \
In the present work both the classic Smagorinsky model and the Germano dynamic procedure \cite{germano:1991} have been employed. 

In Smagorinsky-type subgrid models, the deviatoric part of the subgrid stress tensor $\tau_{ij}$ in (\ref{filteq}) is modelled by a scalar turbulent viscosity $\nu^{\rm sgs}$:
\begin{subequations}\label{eqn:nu_smag}
\begin{align}
 &  \tauij -\frac{1}{3}\tau_{kk} \deltaij = - \frho \nu^{{\rm sgs}} \fSij^d,
\label{eqn:nu_smag:tauij}
 \\
& \nu^{{\rm sgs}} =  C_S^2\Delta^2 |\fS|,
\label{eqn:nu_smag:nu}
\end{align}
\end{subequations}
where $C_S=0.1$ is the Smagorinsky constant, $|\fS|^2 = \dfrac{1}{2}\fSij\fSij$ and $\Delta$ is the filter scale. 
The isotropic part of the subgrid stress tensor can be modelled as in \cite{abba:2015}:
\begin{equation}\label{eqn:taukk_smag}
 \taukk = C_I \frho\Delta^2 |\fS|^2.
\end{equation}
The subgrid temperature flux is set proportional to the resolved temperature gradient:
\begin{equation}\label{eqn:Qj_smag}
 Q_i^{{\rm sgs}} = - \frac{1}{\PR^{{\rm sgs}}} \frho \nusgs \de_i\fT,
\end{equation}
where $\PR^{{\rm sgs}}$ is a subgrid Prandtl number. 
Finally the term $\tau(u_i,u_j,u_k)$ in the subgrid turbulent diffusion flux $J_j^{{\rm sgs}}$ is neglected by analogy with RANS leading to \cite{abba:2015}:
\begin{equation}\label{eqn:Jj_smag}
 J_i^{{\rm sgs}} \approx 2\fu_k\tau_{ik} + \fu_i\taukk.
\end{equation}

In the  Germano dynamic model \cite{germano:1991},  the terms $C_S$ and $C_I$ in the Smagorinsky model are no more chosen \textit{a priori} for the whole domain, but are  computed dynamically as  functions of the resolved field. The deviatoric part of the stress tensor is the same as in the Smagorinsky model:
\begin{equation}
  \tauij -\frac{1}{3}\tau_{kk} \deltaij = - \frho C_S \Delta^2 |\fS| \fSij^d.
\label{eqn:nu_iso:tauij}
\end{equation}
The coefficient $C_S$ is dynamically computed by introducing a test filter operator $\hat{\cdot}$. This operator is linked to the numerical discretization and will be precisely defined in the next section; here it will suffice to point out that, as the filter operator $\bbar{\cdot}$, the test filter is associated to a spatial scale $\hdelta$ (larger than the spatial scale $\Delta$ associated to $\bbar{\cdot}$). A Favre filter denoted with $\breve{\cdot}$ is associated to the test filter through the following Favre decomposition:
\begin{equation}\label{eqn:testfavre_decomp}
 \wh{\rho f} = \wh{\rho} \breve{f}.
\end{equation}
If the test filter $\hat{\cdot}$ is applied to the momentum equation (\ref{filteq-momentum}) we obtain:
\begin{subequations}
\begin{align}
\de_t \left( \wh{\rho} \breve{u}_i \right) + \de_j \left(\wh{\rho} \breve{u}_i \breve{u}_j\right) 
+ \de_i \widehat{p} - \de_j
\widehat{\sigma}_{ij} \nonumber \\
= - \de_j \left( \wh{\tau}_{ij} + \Lij \right),
\label{eq:momentum-test-averages}
\end{align}
\end{subequations}
where
\begin{equation}\label{eqn:leo_qdm}
 \Lij = \wh{\frho\fu_i\fu_j} - \hfrho\hfu_i\hfu_j.
\end{equation}
is the Leonard stress tensor. 
We now assume that the deviatoric part of the Leonard stress tensor can be modelled using an eddy viscosity model:
\begin{equation}
\wh{\tau}^d_{ij} + \Lij^d =
-\hfrho \hdelta^2 |\hfS| C_S \breve{\fS^d}_{rs}
\label{eq:model-test-filter}.
\end{equation}
Substituting (\ref{eqn:nu_iso:tauij}) for $\tau_{ij}^{d}$ and using a least square approach we obtain for the Smagorinsky constant $C_S$ the following expression:
\begin{equation}\label{eq:dynamic-C_S}
 C_S = \dfrac{ \Lij^d \mR_{ij}}{\mR_{kl}\mR_{kl}},
\end{equation}
where
\begin{equation}\label{eq:dynamic-R_kl}
 \mR_{kl} =  \wh{\frho \Delta^2 |\fS| \fS^d_{kl}} -
  \hfrho \hdelta^2 |\hfS| \breve{\fS^d}_{kl}.
\end{equation}
The same dynamic procedure is applied also to the isotropic component of the subgrid stress tensor:
\begin{equation}\label{eqn:taukk_iso}
 \taukk = C_I \frho\Delta^2 |\fS|^2,
\end{equation}
where the $C_I$ coefficient is determined by
\begin{equation}\label{eqn:dynamic-C_I}
 C_I= \dfrac{ \Lkk}{ \hfrho \hdelta^2 |\hfS|^2 - \wh{\frho \Delta^2 |\fS|^2}}.
\end{equation}
A similar approach is proposed also for the subgrid terms in the energy equation. 
For the subgrid heat flux we obtain: 
\begin{equation}\label{eqn:Qj_iso}
Q_i^{{\rm sgs}} = -\frho \Delta^2 |\fS| C_Q \de_i \fT.
\end{equation}
After having applied a dynamic procedure we have:
\begin{equation} \label{eq:dynamic-C_Q}
C_Q = \frac{\Li^Q \mR^Q_i}{\mR^Q_k \mR^Q_k},
\end{equation}
with
\begin{equation}
 \mR^Q_i = \wh{ \frho \Delta^2 |\fS| \de_i \fT} -
 \wh{\frho} \hdelta^2 |\hfS| \de_i \hfT.
\end{equation}
and 
\begin{equation}
\Li^Q = \wh{\frho \fu_i \fT} - \hfrho\hfu_i\hfT,
\label{eq:leonard-temp}
\end{equation}
where $\Li^Q$ is the temperature Leonard flux.
Considering the subgrid turbulent diffusion flux, contrary to what is done in the classic Smagorinsky model, the term $\tau(u_i,u_k,u_k)$ is approximated as a subgrid energy flux which, in turn, can be modelled as a function of the gradient of the resolved kinetic energy:
\begin{equation}\label{eqn:Jj_model_iso}
 \tau(u_i,u_k,u_k) = - \frho \Delta^2 |\fS|
 C_J \de_i\left(\dfrac{1}{2}\fu_k\fu_k\right).
\end{equation}
The value of the coefficient $C_J$ is then determined as:
\begin{equation} \label{eq:iso-CJ}
C_J = \frac{\Li^J \mR_i^J}{\mR_k^J \mR_k^J},
\end{equation}
where
\begin{equation}
 \mR_i^J = \wh{ \frho \Delta^2 |\fS| \de_i\left( \frac{1}{2}\fu_k\fu_k\right)}
- \hfrho \hdelta^2 |\hfS| \de_i \left( \frac{1}{2}\hfu_k\hfu_k\right).
\end{equation}
and $\Li^J$ is the kinetic energy Leonard flux defined as:
\begin{equation}
\Li^J = \hspace{-3pt}\wh{\phantom{\hspace{4pt}}\frho\fu_i\fu_k\fu_k} -
\hfrho\hfu_i\hfu_k\hfu_k.
\label{eq:leonard-ekin}
\end{equation}
It is important to point out that all the dynamic coefficients are averaged over each element in order to avoid numerical instabilities; moreover, since the dynamic model allows backscattering, a clipping procedure analogous to the one introduced in \cite{abba:2015} is applied to ensure that the total dissipation, resulting from both the viscous and the subgrid stresses, is positive.

\section{Numerical method} 
\label{dgmeth}
The equations introduced in section \ref{model}, together with the subgrid scale models of section \ref{turbulence_model}, are spatially discretized by the Discontinuous Galerkin finite elements method. The DG approach is analogous to that described in \cite{giraldo:2008}. In particular the Local Discontinuous Galerkin (LDG) method is chosen for the approximation of the second order viscous terms (see \cite{arnold:2002}, \cite{bassi:1997},  \cite{castillo:2000}, \cite{cockburn:1998}). 
In the LDG method, the non-dimensional system of Navier-Stokes equations (\ref{filteq}) is rewritten introducing an auxiliary variable $\bmG$, so that
\begin{eqnarray}
\label{eq:csv_auxvar}
\de_t \bU + \Div \bF^{{\rm c}}(\bU) &=&  \Div \bF^{{\rm v}}(\bU,\bmG) \nonumber\\
 &-&\Div \bF^{{\rm sgs}}(\bU,\bmG) + \bS   \\
 \bmG &-& \nabla{\boldsymbol \varphi} = 0,\nonumber
\end{eqnarray}
where $\bU=[\frho\,,\frho\wt{\bu}^T,\frho\fe ]^T$ are the prognostic
variables, ${\boldsymbol \varphi} = [\wt{\bu}^T,\fT]^T$ are the variables
whose gradients enter the viscous fluxes~(\ref{eq:constitutive-Favre}),
as well as the turbulent ones  and $\bS$ represents the source terms.
The fluxes in (\ref{eq:csv_auxvar}) are written in the following compact form:
\[
\bF^{{\rm c}} = \left[
\begin{array}{c}
 \frho\wt{\bu} \\
 \frho\wt{\bu}\otimes\wt{\bu} +
 \bbar{p}\mathcal{I} \\
 \frho\wt{h}\wt{\bu}
\end{array}
\right],
\]
\[
\bF^{{\rm v}} = \left[
\begin{array}{c}
 0 \\
 \wt{\sigma} \\
  \wt{\bu}^T \wt{\sigma}
 -  \wt{\bq}
\end{array}
\right]
\]
and
\[
\bF^{{\rm sgs}} = \left[
\begin{array}{c}
 0 \\
 \tau \\
 \frac{1}{(\gamma-1)\MA^2} \mathbf{Q}^{{\rm sgs}}
 +\frac{1}{2}\left( 
 \mathbf{J}^{{\rm sgs}} - \tau_{kk} \wt{\bu}
 \right)
\end{array}
\right],
\]
\[
\bS = \left[
\begin{array}{c}
 0 \\
 \frho \mathbf{f} \\
\frho \mathbf{f} \cdot \wt{\bu}
\end{array}
\right].\]
Here, $\tau$, $\mathbf{Q}^{{\rm sgs}}$ and $\mathbf{J}^{{\rm sgs}}$
are given by~(\ref{eqn:nu_smag}), (\ref{eqn:Qj_smag})
and~(\ref{eqn:Jj_smag}), respectively, for the Smagorinsky model while they are given by~(\ref{eqn:nu_iso:tauij}), (\ref{eqn:Qj_iso})
and~(\ref{eqn:Jj_sgs2}) together with~(\ref{eqn:Jj_model_iso}) for the dynamic model. 

To define the space discretization, a tessellation $\mT_h$ of $\Omega$ into tetrahedral elements $K$ such that $\Omega =
\bigcup_{K\in\mT_h} K$ and $K\cap K'=\emptyset$ is introduced and the finite element space is defined as:
\begin{equation}\label{eqn:mV_def}
\mV_h = \left\{ v_h \in L^2(\Omega): v_h|_K \in \mathbb{P}^q(K), \,
\forall K\in\mT_h \right\},
\end{equation}
where $q$ is a nonnegative integer and $\mathbb{P}^q(K)$ denotes the
space of polynomial functions of total degree at most $q$ on $K$. 
For
each element, the outward unit normal on $\partial K$ will be denoted
by $\bn_{\partial K}$. Given $d$ the dimension of the problem the numerical solution is now defined as
$(\bU_h,\bmG_h)\in(\,(\mV_h)^{(2+d)}\,,\,(\mV_h)^{4\times d}\,)$ such that,
$\forall K\in\mT_h$, $\forall v_h\in\mV_h$, $\forall
\br_h\in(\mV_h)^d$,

\begin{subequations}
\label{eq:DG-space-discretized}
\begin{align}
\displaystyle
\frac{d}{dt}\int_K \bU_h v_h\,d\bx
& \displaystyle
- \int_K \bF(\bU_h,\bmG_h)\cdot\nabla v_h\, d\bx
\\[3mm]
& \displaystyle
+ \int_{\partial K} \bF^*(\bU_h,\bmG_h)\cdot \bn_{\partial K} v_h\,
d\sigma
= \int_K \bS v_h \,d\bx,\nonumber 
\\[3mm] \displaystyle
\int_K \bmG_h \cdot \br_h \,d\bx
& \displaystyle
+ \int_K {\boldsymbol \varphi_h}\nabla\cdot\br_h\, d\bx
\\[3mm]
& \displaystyle
- \int_{\partial K} {\boldsymbol \varphi}^* \bn_{\partial
K}\cdot\br_h \, d\sigma = 0, \nonumber 
\end{align}
\end{subequations}
where $\bU_h=\left[ \rho_h\,,\rho_h\bu_h\,,\rho_he_h \right]^T$,
${\boldsymbol \varphi}_h=\left[ \bu_h\,,T_h \right]^T$, $\bF =
\bF^{{\rm c}}-\bF^{{\rm v}}+\bF^{{\rm sgs}}$, and $\bF^*$,
${\boldsymbol \varphi}^*$ denote the so-called
 numerical fluxes. The numerical fluxes are responsible for the coupling among different elements. In this work
 we have tested   a) the Rusanov flux, b) a modified version of the Rusanov flux appropriate for
 low Mach number flows, employing   the velocity of the fluid as upwinding velocity c) the exact Godunov Riemann solver (implemented as in \cite{gottlieb:1988}). In order to avoid some numerical difficulties in simulations with
 small density differences (see the discussion in section  \ref{numfluxes}),  
 the last one has been extensively employed in the simulations for $\bF^*$. The centered flux is employed for ${\boldsymbol\varphi}^*$. 
On each element, the unknowns are expressed in terms of an orthogonal polynomial basis, yielding what is
commonly called a modal DG formulation. All the integrals are evaluated using quadrature formulae
from~\cite{cools:2003}, which are exact for polynomial orders up to $2q$. This results in a diagonal mass matrix in the time derivative term of~(\ref{eq:DG-space-discretized}) and simplifies the computation of $L^2$ projections to be introduced shortly in connection with the LES filters. 

In the following the filter operators $\bbar{\cdot}$ and $\wh{\cdot}$, introduced in section~\ref{turbulence_model}, will be explicitly defined in the context of the DG finite elements method. In particular the filters operators are defined in terms of an $L^2$ projection, as suggested e.g. in \cite{collis:2002b},  \cite{collis:2002a}, \cite{vanderbos:2007}. 
Given a subspace $\mV\subset L^2(\Omega)$, let
$\Pi_{\mV}:L^2(\Omega)\to\mV$ be the associated projector defined by
\[
\int_\Omega \Pi_{\mV}u\,v\, d\bx =
\int_\Omega u\,v\, d\bx, \qquad \forall u,v \in\mV,
\]
where the integrals are evaluated with the same quadrature rule used
in~(\ref{eq:DG-space-discretized}). For $v\in L^2(\Omega)$, the filter
$\bbar{\cdot}$ is now defined by
\begin{equation}
\bbar{v} = \Pi_{\mV_h}v.
\label{eq:filter-bar}
\end{equation}
Notice that the application of this filter is built in the
discretization process and equivalent to it. Therefore,
once the discretization of equations (\ref{eq:csv_auxvar}) has
been performed, only $\bbar{\cdot}$ filtered
quantities are computed by the model.
To define the test filter, we  then introduce
\begin{equation}\label{eqn:mVhat_def}
\wh{\mV}_h = \left\{ v_h \in L^2(\Omega): v_h|_K \in
\mathbb{P}^{\wh{q}}(K), \, \forall K\in\mT_h \right\},
\end{equation}
where $0\leq\wh{q}<q$, and we let, for $v\in L^2(\Omega)$,
\begin{equation}
\wh{v} = \Pi_{\wh{\mV}_h}v.
\label{eq:filter-hat}
\end{equation}
By our previous identification
of the $\bbar{\cdot}$  filter and the discretization,
the quantities $\frho$, $\frho\wt{\bu}$ and $\frho\fe$
can be identified with
  $\rho_h$, $\rho_h \bu_h $ and $\rho_he_h,$ respectively.
  Therefore, they belong to $\mV_h,$ for which an orthogonal
  basis is employed by the numerical method.
  As a result, the computation of $\wh{\rho_h}$, $\wh{\rho_h\bu_h}$
and $\wh{\rho_he_h}$ is straightforward and reduces to zeroing the
last coefficients in the local expansion.
Assuming that the analytic solution is defined in some infinite
dimensional subspace of $L^2$, heuristically, $\mV_h\subset L^2$
is associated to the scales which are represented by the model, while
$\wh{\mV}_h\subset\mV_h\subset L^2$  is associated to
the spatial scales  well resolved by the numerical approximation. 

The filtering operations (\ref{eq:filter-bar}) and (\ref{eq:filter-hat}) are realized by imposing pointwise the conditions (\ref{ffilter_collective}), (\ref{eqn:tauij_sgs}) and (\ref{eq:constitutive-Favre}). The Leonard stress tensors (\ref{eqn:leo_qdm}), (\ref{eq:leonard-temp}) and (\ref{eq:leonard-ekin}) are computed using (\ref{eq:filter-hat}) with the quadrature rules given in \cite{cools:2003}. 
Notice that the filters defined by $L^2$ projections do not commute with differential operators. However,  in this work we neglect the commutation errors.

The spatial scales $\Delta$ and $\hdelta$ associated with the two filters~(\ref{eq:filter-bar}) and~(\ref{eq:filter-hat}) can be computed by dividing the element diameter by the cubic root (or the square root in two dimensions) of the number of degrees of freedom of $\mathbb{P}^q(K)$, for $\Delta$,
and $\mathbb{P}^{\wh{q}}(K)$, for $\hdelta$; the filter scales are, as a consequence, piecewise constant functions in space.   Finally, time integration was performed by  a 4 stage explicit Runge-Kutta method.  

All the computations were performed using the implementation of the above described method provided in the  finite
element library {\tt FEMilaro} \cite{femilaro}. This tool exploits
 modern FORTRAN/MPI  features, aims at providing a flexible
environment for the development and testing of new finite element
formulations and is publicly available under GPL license.

\section{The lock exchange benchmark}
\label{lock}
\begin{figure}
\centering
\includegraphics[width=0.8\textwidth]{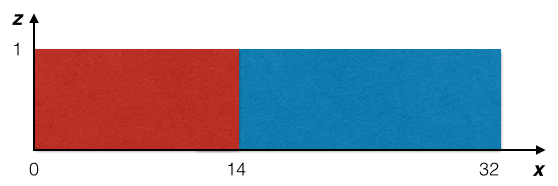}
\caption{Schematic representation of the initial datum for the lock-exchange configuration.}
\label{fig:lock_exchange_domain}
\end{figure}

The lock-exchange configuration used to assess the capability of the DG-LES model
is represented in Figure \ref{fig:lock_exchange_domain}. The domain length is  $L=32$ and its height is $H=1.$ A membrane at $x_0=14$ initially divides the rectangular container into two compartments. In our case, the two chambers are filled with the same fluid at different densities on the two sides of the membrane (higher density on the left and lower density on the right). Upon the removal of the membrane, the dense front moves rightward along the lower boundary, while the light front propagates leftward along the upper boundary. 

The quantities   chosen as reference physical quantities for nondimensionalization of the compressible NS equations are the height of the channel $H^*$, the larger of the two fluid densities $\rho_1^*$ and the buoyancy velocity $u_b^*=\sqrt{{g'}^* H^*}$. Notice that ${g'}^*$ is the reduced gravity, which is computed as:
\begin{equation}
{g'}^* = g^* \frac{\rho^*_1-\rho^*_2}{\rho^*_1},
\label{reduced_gravity}
\end{equation}
where  $g^*$ denotes the gravity acceleration and $\rho^*_2$ the lower density.  
Some relationships exist between the nondimensional numbers $Fr$ and $\MA$ and the characteristic quantities of the test case. 
For the Froude number $Fr,$ the following equation is valid:
\begin{equation}
\frac{1}{Fr^2} = \frac{g^* H^*}{{u_b^*}^2} = \frac{g^*}{g^* \frac{\rho_1^*-\rho_2^*}{\rho_1^*}}  = \frac{1}{1-\gamma_r},
\label{froude}
\end{equation}
where $\gamma_r=\rho_2/\rho_1$ is the ratio between densities. 
If we write the definition of the Mach number as the ratio between the characteristic velocity (i.e., the buoyancy velocity) and the sound speed we obtain:
\begin{equation}
{\MA}^2 = \frac{{u_b^*}^2}{\frac{\gamma p^*}{\rho_1^*}} = \frac{g^* H^* (\rho_1^*-\rho_2^*)}{\gamma p^*}.
\end{equation}
 If we write the dimensional pressure $p^*=p p_r = p \rho_1^* {u_b^*}^2$ as the product of the 
 nondimensional pressure multiplied by the reference pressure and we simplify, we obtain for the Mach number the following relationship:
\begin{equation}
\MA = \frac{1}{\sqrt{\gamma p}},
\label{mach}
\end{equation}
where $\gamma$ is the ratio between the specific heats. 
Concerning the initial conditions, even though the DG finite element method would be able to manage a discontinuous initial datum, in order to better reproduce the results obtained in literature,
 the initial density profile is smoothed out as in \cite{birman:2005}: 
\begin{equation}
\rho_0(x) = \frac{\gamma_r+1}{2} - \frac{1-\gamma_r}{2} {\rm erf}
\left( \frac{x-x_0}{\sqrt{Re}} \right),
\label{init_dens_prof}
\end{equation}
where $x$ denotes the horizontal coordinate.
Since we are considering the compressible Navier-Stokes equations, it is necessary to specify the initial conditions also for pressure and temperature. 
The initial pressure value at the top of the domain is computed by using equation (\ref{mach}) as $p_{in}^{top}=1/(\gamma \MA^2)$. The initial pressure value in the whole domain is  then computed assuming an hydrostatic pressure profile:
\begin{equation}
p_{in} = p_{in}^{top} + \rho(x)(1-z)/{Fr}^2,
\label{init_press_prof}
\end{equation} 
where   $z$ denotes the vertical coordinate.
The initial datum for temperature is derived starting from density and pressure and using the equation of state. 
Since our aim is to reproduce incompressible results, a Mach number of 0.008 has been chosen. 
In order to achieve effective comparisons with \cite{birman:2005}, two-dimensional simulations with slip boundary conditions have been performed.  The performance of our model has been evaluated by comparing
density contours and, more quantitatively, different kinds of energy budgets. 
In particular, we have computed the time evolution of the normalized potential, kinetic and dissipated energies integrated over the whole domain. The potential energy is:
\begin{equation}
E_p(t) = \int_{\Omega} \frac{1}{Fr^2} \rho z dV.
\end{equation}
The kinetic energy is:
\begin{equation}
E_k(t) = \int_{\Omega} \frac{1}{2} \rho u_i u_i dV,
\end{equation}
while the time evolution of the dissipated energy is computed by solving the following equation: 
\begin{equation}
\frac{dE_d}{dt} = \int_\Omega \biggl\{ \mu \biggl[ \frac{1}{2}(\partial_j u_i + \partial_i u_j)^2-\frac{2}{3}(\nabla \cdot \textbf{u})^2\biggr] \biggr\} dV.
\end{equation}
As proposed in \cite{birman:2005} the quantity $E_{amb}$ is computed as:
\begin{equation}
E_{amb} = \frac{1}{1-\gamma_r}\int_\Omega \gamma_r z dV.
\end{equation}
From the initial potential energy, by subtracting $E_{amb}$, 
we obtain $E_{p0}^a$ (defined in \cite{birman:2005} as initial available potential energy):
\begin{equation}
E_{p0}^a = E_{p0}-E_{amb}
\end{equation}
and the available potential energy at a generic time $t$: 
\begin{equation}
E_p^a(t) = E_p(t)-E_{amb}.
\end{equation}
We then normalize each contribution to the overall energy budget with $E_{p0}^a$ and we obtain the normalized potential energy $E_p^n(t)=E_p^a(t)/E_{p0}^a$, the normalized kinetic energy $E_k^n(t)=E_k(t)/E_{p0}^a$ and the normalized dissipated energy $E_d^n(t) = E_d(t)/E_{p0}^a$. 
The second energy budget taken into consideration provides the evaluation of the energies dissipated by the dense and the light front separately, as suggested in \cite{birman:2005}. The energy dissipated by the light front is computed using the following equation:
\begin{equation}
\frac{dE_d^{light}(t)}{dt} = \int_{0}^{x_0} \int_0^{1} \mu \biggl[ \frac{1}{2}(\partial_j u_i + \partial_i u_j)^2-\frac{2}{3}(\nabla \cdot \textbf{u})^2\biggr] \biggr\} dz dx,
\end{equation}
while the energy dissipated by the dense front is:
\begin{equation}
\frac{dE_d^{dense}(t)}{dt} = \int_{x_0}^{L} \int_0^{1} \mu \biggl[ \frac{1}{2}(\partial_j u_i + \partial_i u_j)^2-\frac{2}{3}(\nabla \cdot \textbf{u})^2\biggr] \biggr\} dz dx.
\end{equation}
where $x_0$ is the abscissa of the initial discontinuity.

\subsection{Direct Numerical Simulations}

For the fully resolved simulations,
a grid composed by approximately  $37000$ elements has been employed. The polynomial degree has been set equal to $4,$ while the polynomial degree associated to the test filter operation was 
taken to be $2.$ Notice that, with these choices, the total number of degrees of freedom is of the same order of magnitude as that in \cite{birman:2005}. 

\begin{figure}[]
\centering
\begin{subfigure}[]{
      \label{fig:density_contours_a}
\hspace{-7mm}
      \includegraphics[width=1.1\textwidth]{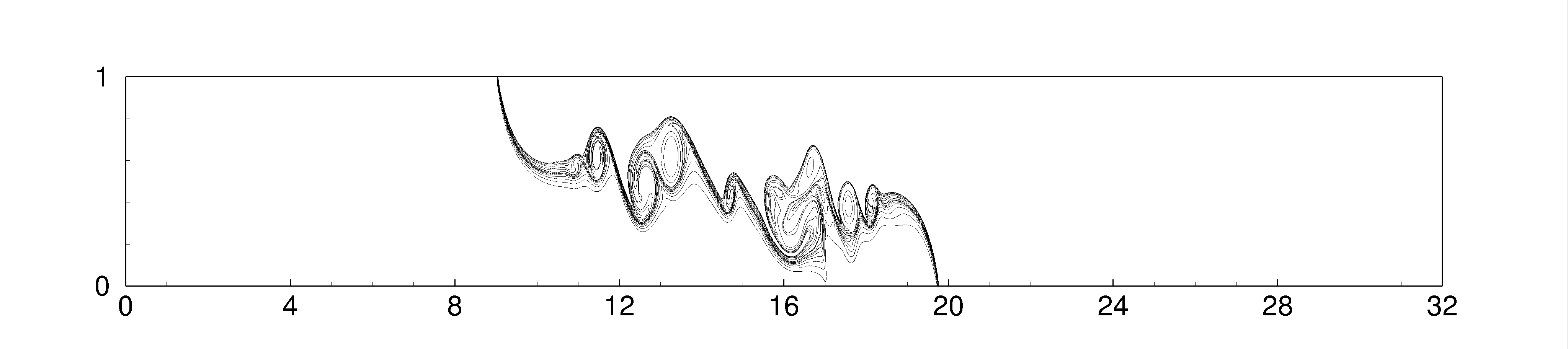}}
\end{subfigure}
\begin{subfigure}[]{
      \label{fig:density_contours_b}
      \includegraphics[width=1.0\textwidth]{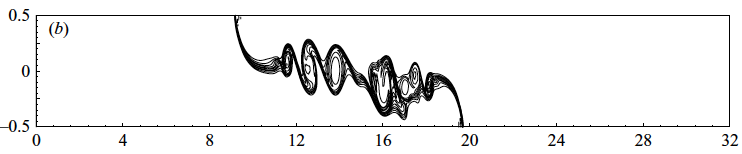}}
\end{subfigure} 
\begin{subfigure}[]{
        \label{fig:density_contours_c}
\hspace{-7mm}
       \includegraphics[width=1.1\textwidth]{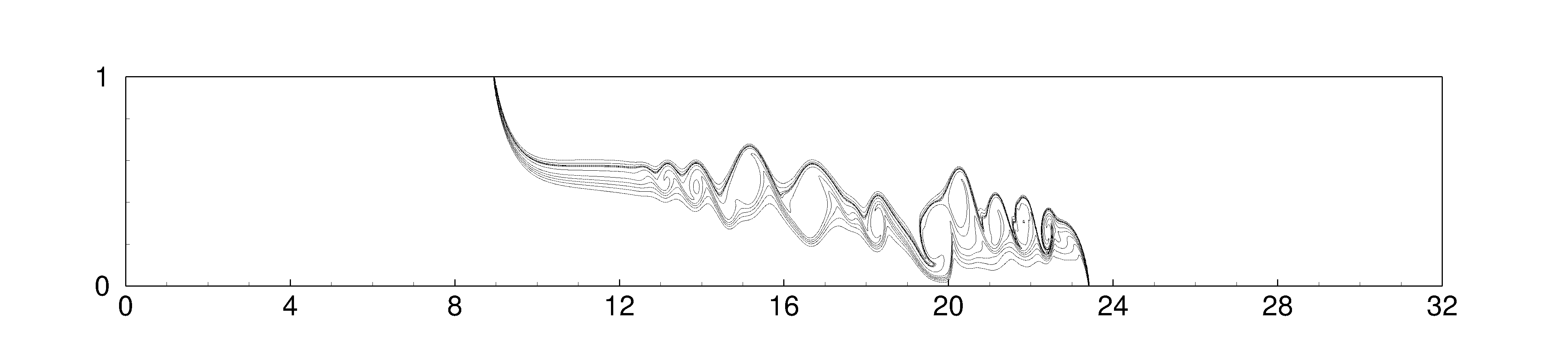}}
\end{subfigure}
\begin{subfigure}[]{
      \label{fig:density_contours_d}
      \includegraphics[width=1.0\textwidth]{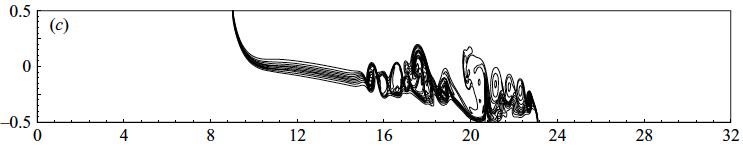}}
\end{subfigure}
\caption{Density contours at $t=10$ for $Re=4000$ (DNS) (a) DG simulation, $\gamma_r=0.7$ (b)\cite{birman:2005}, $\gamma_r=0.7$ (c) DG simulation, $\gamma_r=0.2$ (d) \cite{birman:2005}, $\gamma_r=0.2$}
\label{fig:density_contours}
\end{figure}

In Figure \ref{fig:density_contours} we compare the density contours obtained from our compressible NS simulations with the corresponding density contours obtained in \cite{birman:2005}, for density ratios $\gamma_r=0.2,0.7$, at $t=10$ and for $Re=4000$. For both density ratios the results are quite similar, both in terms of the position of the front and of the number and appearance of the  Kelvin-Helmoltz billows. In particular, considering the lower density ratio $\gamma_r=0.2$ (Figures \ref{fig:density_contours_c}-\ref{fig:density_contours_d}) we notice that, as in \cite{birman:2005}, the dense front presents a considerably lower height and that, by $t=10$, it has propagated further with respect to the light front. Moreover, by decreasing the density ratio, we observe that the vortical structures appear  to be confined to the region near the dense front. According to the explanation in \cite{birman:2005}, 
this is because across the dense current the velocity difference and the shear are larger, which has destabilizing effect. 

\begin{figure}
\hspace{-7mm}
\includegraphics[width=1.1\textwidth]{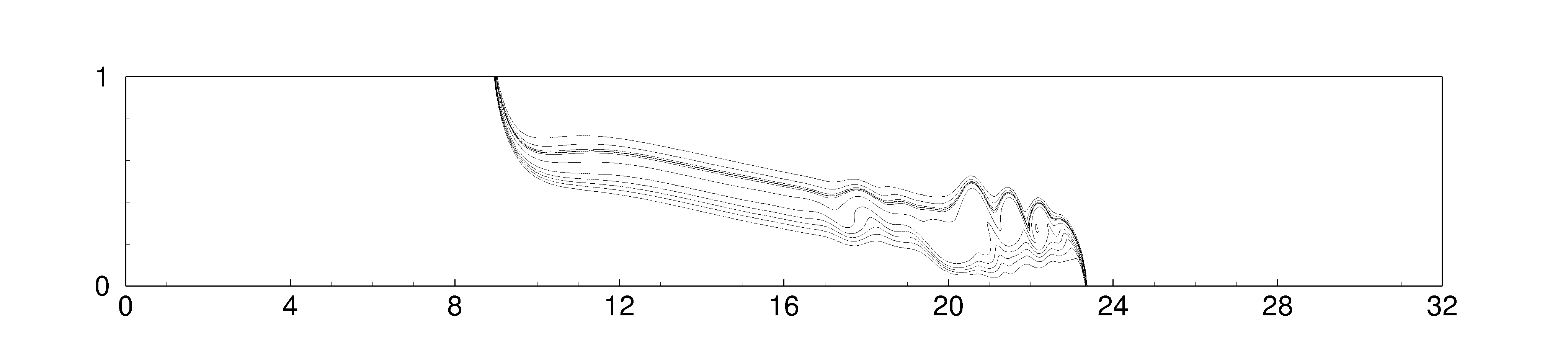}
\caption{Density contours at $t=10$ for $\gamma_r=0.2$ and $Re=4000$ (DNS) with constant dynamic viscosity $\mu=1/Re$.}
\label{fig:density_constant_mu}
\end{figure}

\begin{figure}
\centering
\includegraphics[width=0.8\textwidth]{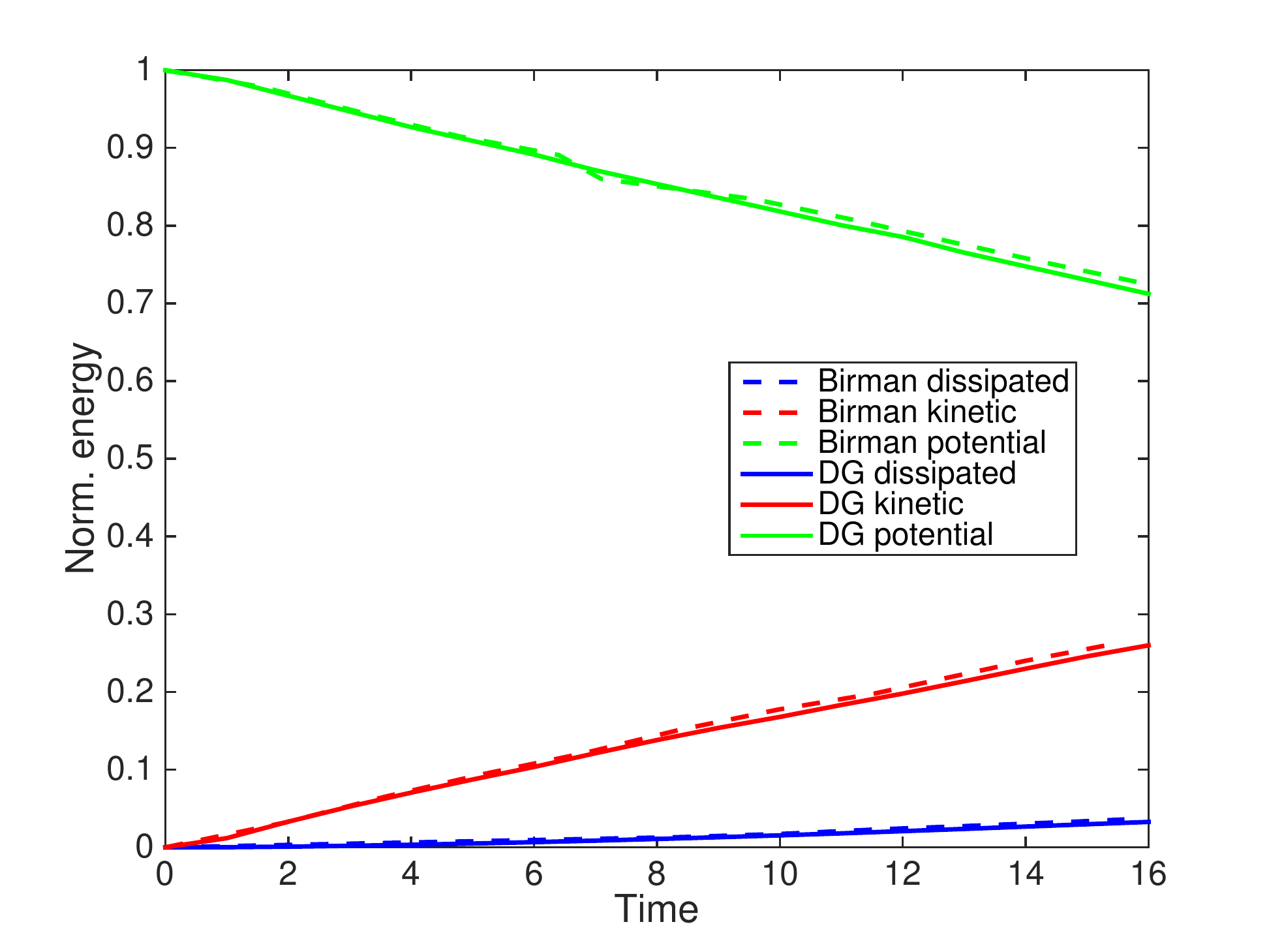}
\caption{Time evolution of the normalized potential energy $E_p^n$ (green), normalized kinetic energy $E_k^n$ (red) and normalized dissipated energy $E_d^n$ (blue) for $\gamma_r=0.4$ and $Re=4000$ (DNS). Continuous lines: DG simulation. Dashed lines: \cite{birman:2005} simulation.}
\label{fig:energy_budget1}
\end{figure}

\begin{figure}
\centering
\includegraphics[width=0.8\textwidth]{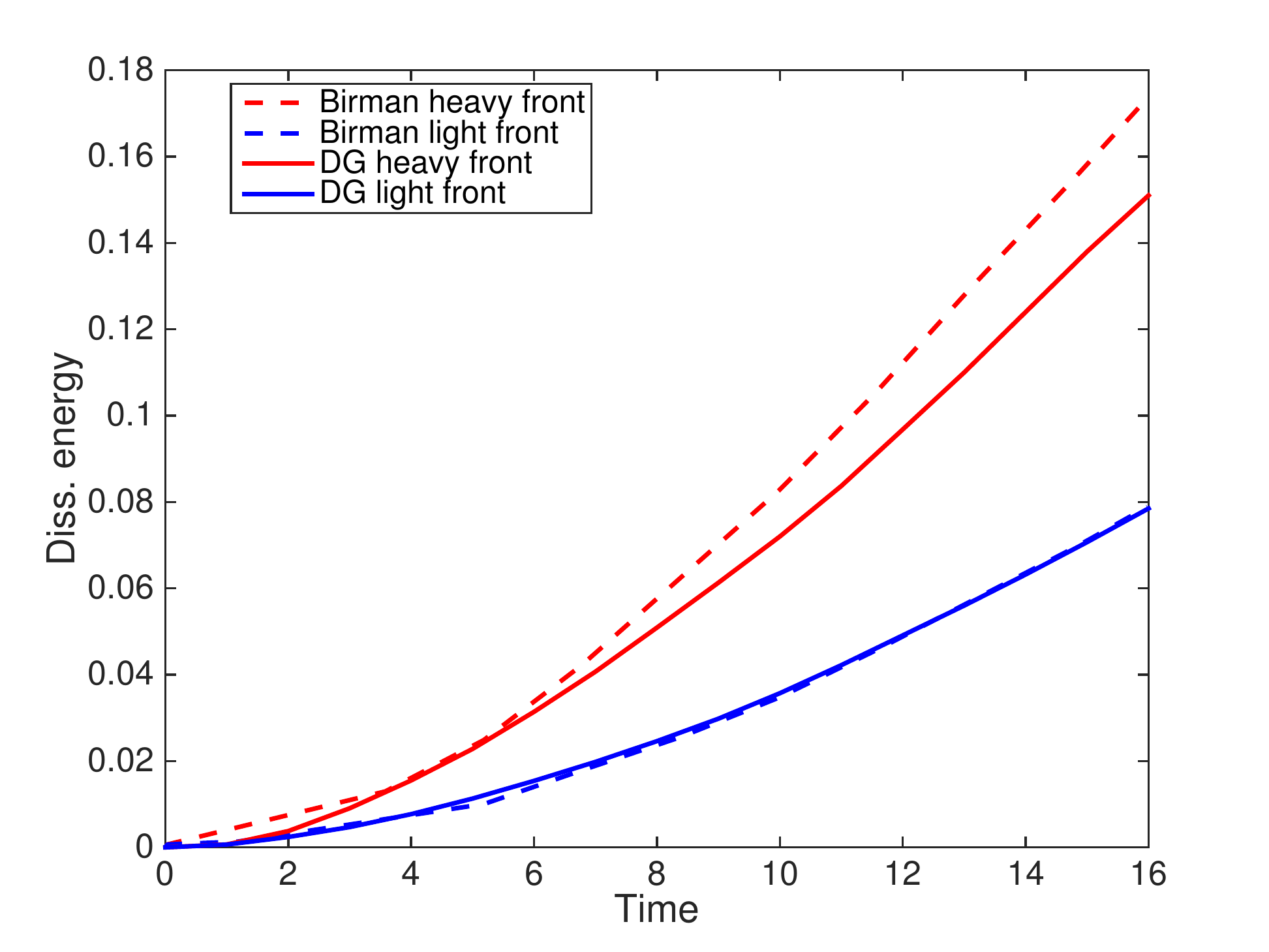}
\caption{Time evolution of the energies dissipated by the dense (red) and light (blue) fronts for $\gamma_r=0.4$ and $Re=4000$ (DNS). Continuous lines: DG simulation. Dashed lines: \cite{birman:2005} simulation.}
\label{fig:energy_budget2}
\end{figure}

In the case of $\gamma_r=0.2$ we notice however some discrepancies between our results and those of \cite{birman:2005}. The length of the area of the density current interested by the presence of Kelvin-Helmoltz billows is more extended in our case: the first vortex is located approximately at $x=13$ in our simulation (see Figure \ref{fig:density_contours_c}) while its position is $x=15$ in \cite{birman:2005} (Figure \ref{fig:density_contours_d}). 
Another important feature which is correctly reproduced by the DG simulations is the behaviour of the density current in the presence of constant dynamic viscosity (see Figure \ref{fig:density_constant_mu}). In this situation the formation of vortexes is limited to a smaller region close to the dense front (as highlighted in \cite{birman:2005}). 
Considering energy budgets, in Figure \ref{fig:energy_budget1} we present the time evolution of the normalized potential energy $E_p^n(t)$, the normalized kinetic energy $E_k^n(t)$ and the normalized dissipated energy $E_d^n(t)$ up to $t=10$, for $Re=4000 $ and density ratio $\gamma_r=0.4$. We notice that the time evolution of the different  energy budgets reported in \cite{birman:2005} (dashed lines) is very well captured by the DG simulation (continuous lines). 
In Figure \ref{fig:energy_budget2} we show the time evolution of the energies dissipated  by the light and the dense front, respectively. Both in \cite{birman:2005} (dashed lines) and in DG our simulation (continuous lines), the dense front (red lines) is more dissipative with respect to the light front (blue lines). However, a slight underestimation of the energy dissipated by the dense front is present in the DG simulation. 
\subsection{Large Eddy Simulations}
Besides simulations at DNS resolution, also some  Large Eddy Simulation experiments have been
carried out in  the lock-exchange test-case, using a density ratio $\gamma_r=0.7$ and Reynolds number $\RE=40000.$ These experiments have been performed employing the same computational grid and the same polynomial degree as in the DNS at $Re=4000$. 
Both the classic Smagorinsky model and the Germano dynamic model have been considered, together with an under-resolved DNS obtained without employing any model. For this case, the simulation could be completed
without having to add any artificial diffusion.
 
\begin{figure}
\centering
\includegraphics[width=0.8\textwidth]{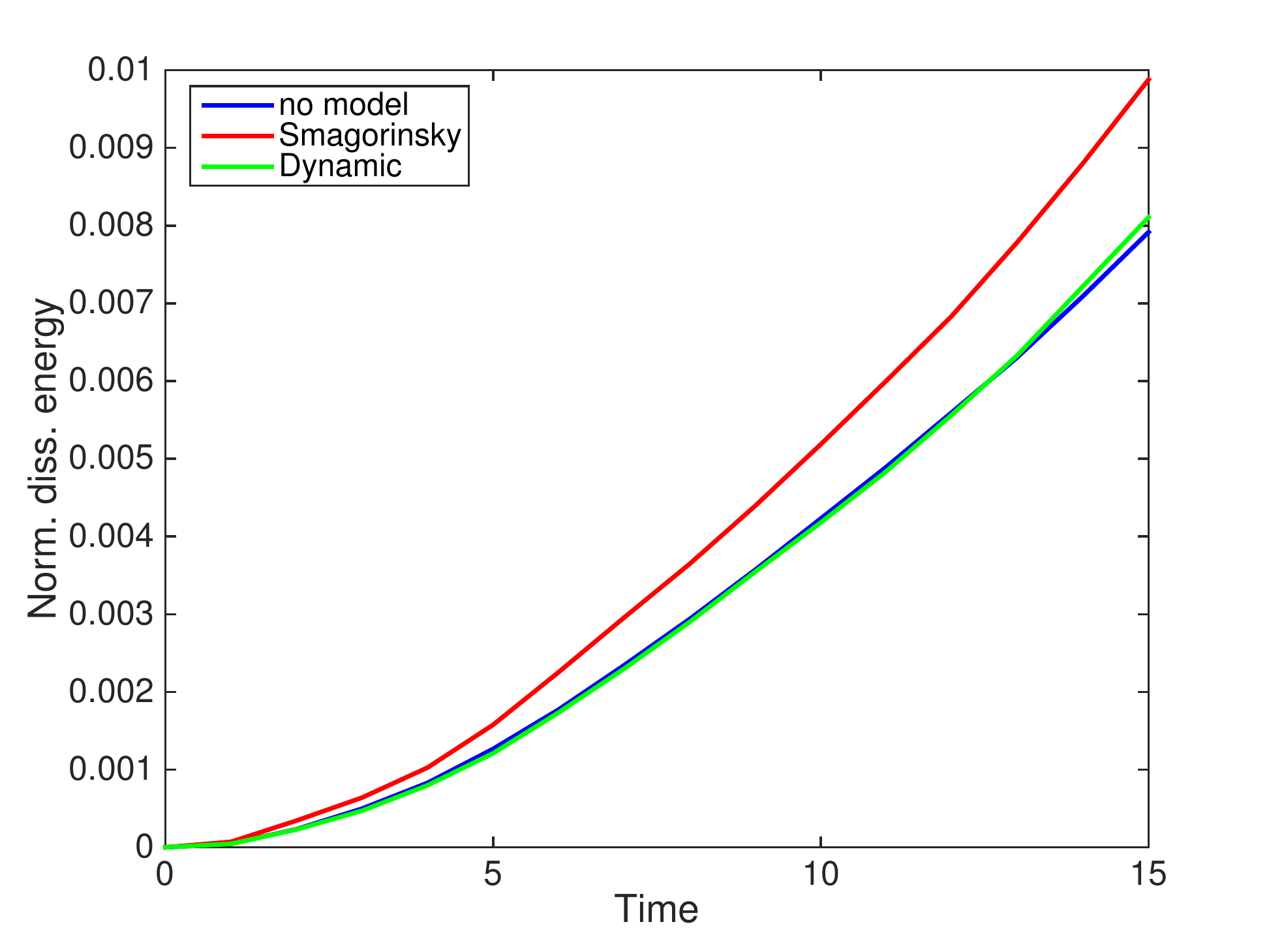}
\caption{Time evolution of the normalized energies dissipated using no model (blue), the Smagorinsky model (red), the dynamic model (green) for $\gamma_r=0.7$ and $\RE=40000$ (LES).}
\label{fig:energy_diss_turb}
\end{figure}

\begin{figure}[]
\centering
\begin{subfigure}[]{
       \label{fig:vorticity_a}
      \includegraphics[width=0.8\textwidth]{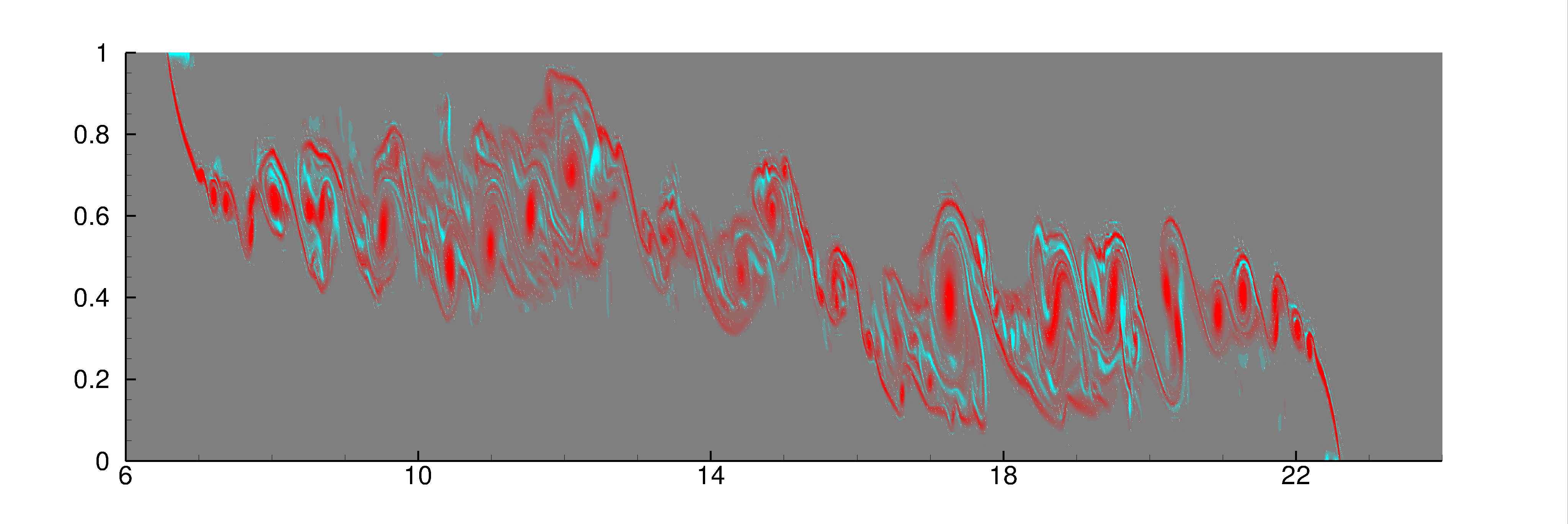}}
\end{subfigure}
\begin{subfigure}[]{
      \label{fig:vorticity_b}
      \includegraphics[width=0.8\textwidth]{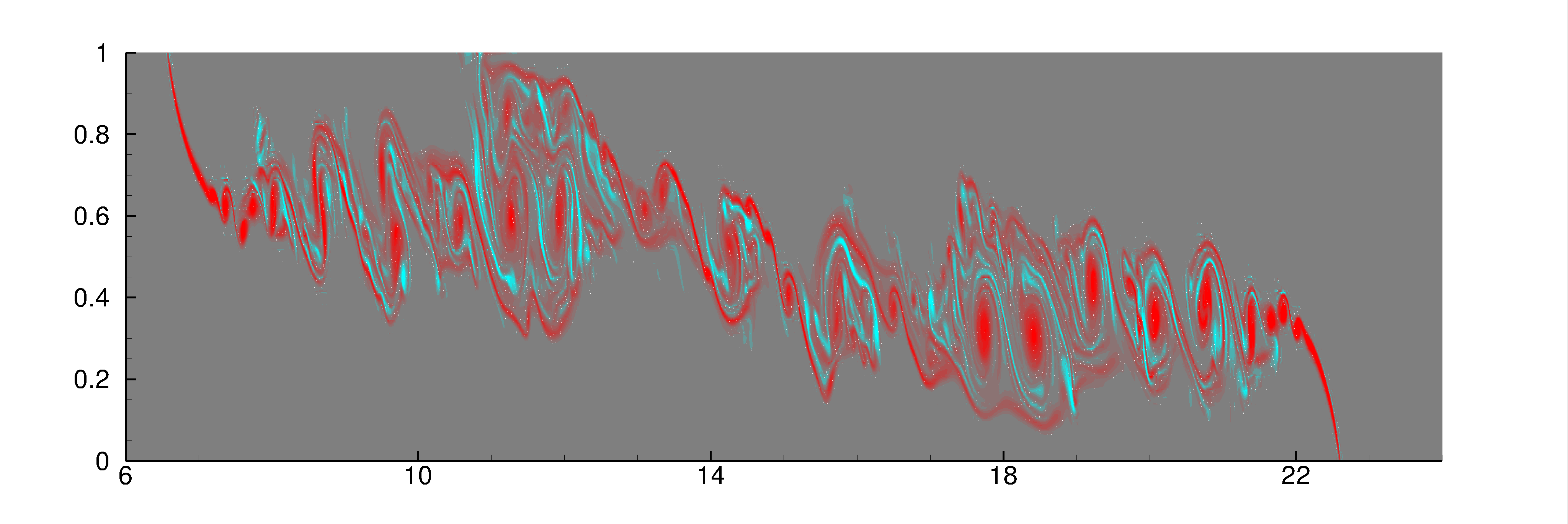}}
\end{subfigure} 
\begin{subfigure}[]{
      \label{fig:vorticity_c}
      \includegraphics[width=0.8\textwidth]{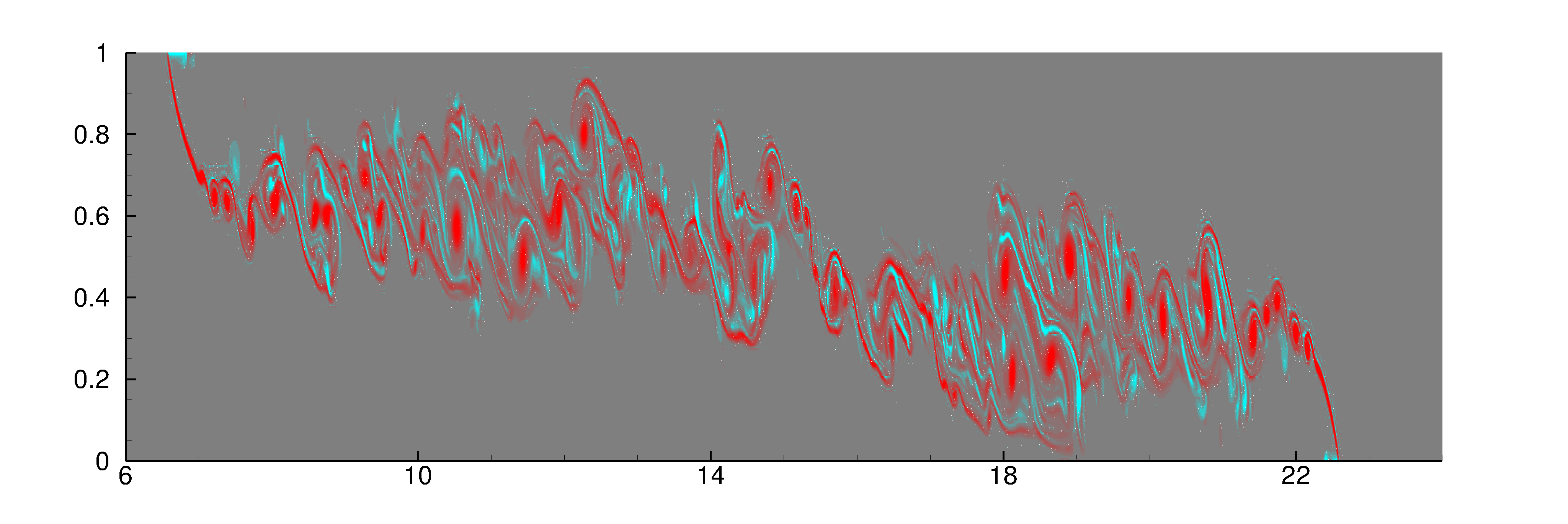}}
\end{subfigure}
\caption{Vorticity field obtained for $\gamma_r=0.7$ and $\RE=40000$ (LES) at $t=15$. Red: positive values. Blue: negative values. (a) No model, value range $[-63,77]$. (b) Smagorinsky model, value range $[-42,54]$. (c) Dynamic model, value range $[-74,85]$.}
\label{fig:vorticity}
\end{figure}

\begin{figure}
\centering
\includegraphics[width=0.8\textwidth]{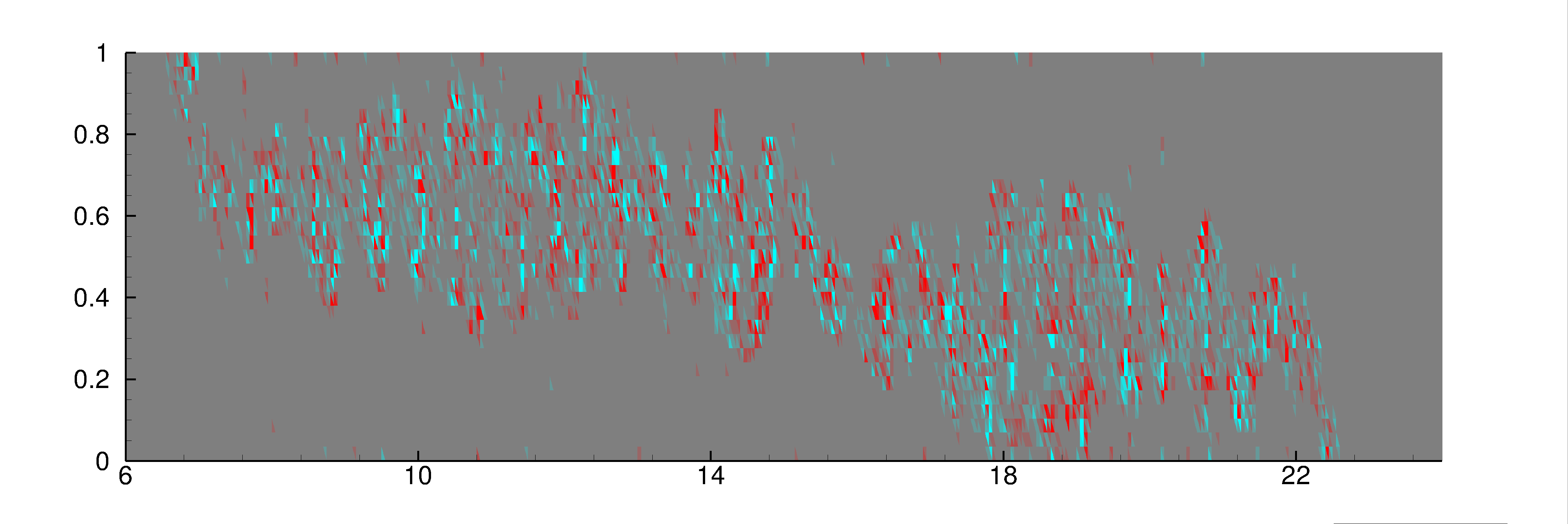}
\caption{Dynamic constant $C_s$ for $\gamma_r=0.7$ and $\RE=40000$ (LES) at $t=15$. Value range $[-0.08,0.1]$. Red: positive values. Blue: negative values.}
\label{fig:dyn_const}
\end{figure}

In Figure \ref{fig:energy_diss_turb} we show the  dissipated energy as a function of time for the different models. We notice that the Smagorinsky model (red line) presents a much more dissipative behaviour with respect to the no-model (blue line) and the dynamic model (green line) simulations.
This behaviour of the Smagorinsky model is confirmed by the vorticity field at a fixed instant of time $t=15$ (see Figure \ref{fig:vorticity}). The vorticity field provided by the Smagorinsky model (Figure \ref{fig:vorticity_b}) is smoother and characterized by less vortical structures and lower vorticity peak values.
As regard as the comparison between no-model and dynamic model we notice that, even if the dissipated energy does not evidence significant differences (as can be easily seen in Figure \ref{fig:energy_diss_turb} observing the green and blue lines), a careful examination of Figure \ref{fig:vorticity} allows to observe that the vorticity field obtained with the dynamic model (Figure \ref{fig:vorticity_c}) presents finer structures and higher vorticity peak values with respect to the no model case (Figure \ref{fig:vorticity_a}), suggesting that backscatter probably plays a quite important role when employing the dynamic model. 
This fact is confirmed also by considering the values of the dynamic constant $C_s$ at the same instant $t=15$ represented in Figure \ref{fig:dyn_const}: there are indeed many elements where $C_s$ assumes negative values (indicated in blue) smaller than $-0.02$, with peak negative values of $-0.08$. These values are not negligible if compared to the value assumed by the Smagorinsky constant in the static Smagorinsky model which is $C_s =0.1$. 

Notice that approximately $13500$ CPU hours have been necessary to carry out the DNS at $\RE=4000$ till $t=15$ and that the same number of CPU hours has been employed for the under-resolved DNS at $\RE = 40000$.
On the other hand, the Large Eddy Simulations at $\RE=40000$ with the Smagorinsky model and with the dynamic procedure have required $20000$ and $22000$ CPU hours, respectively. 
\subsection{Impact of the  numerical flux choice}
\label{numfluxes}
\begin{figure}[]
\centering
\begin{subfigure}[]{
      \label{fig:numflux_a}
      \includegraphics[width=0.8\textwidth]{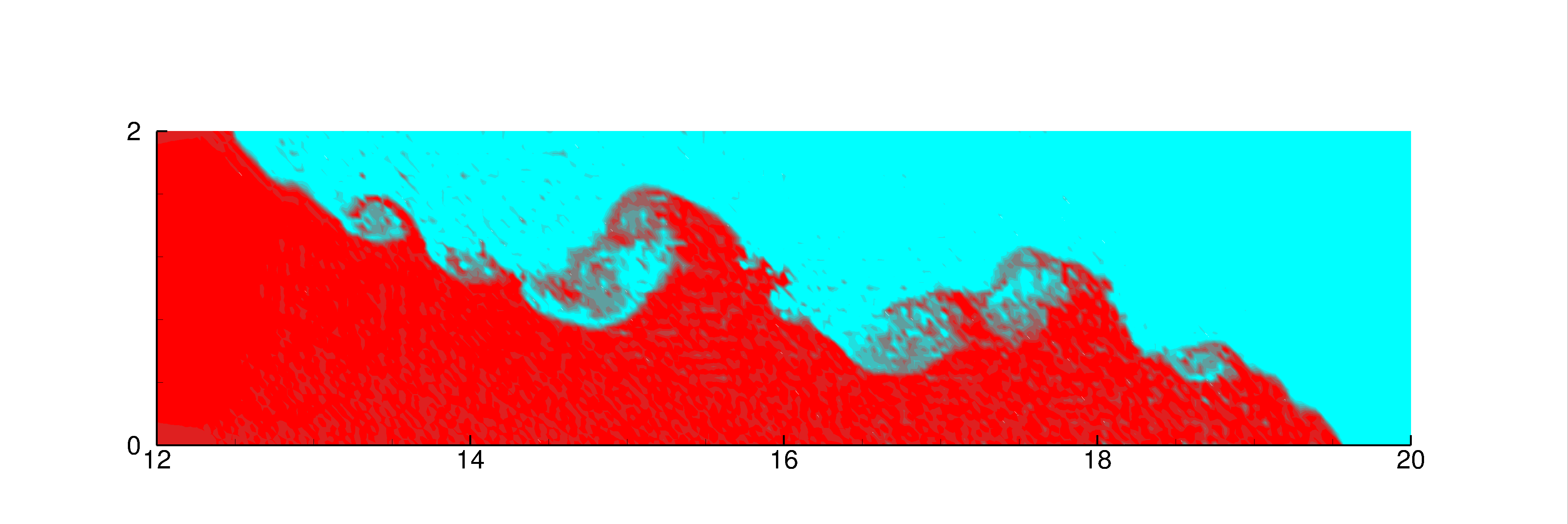}}
\end{subfigure}
\begin{subfigure}[]{
     \label{fig:numflux_b}
\vspace{-14mm}
      \includegraphics[width=0.8\textwidth]{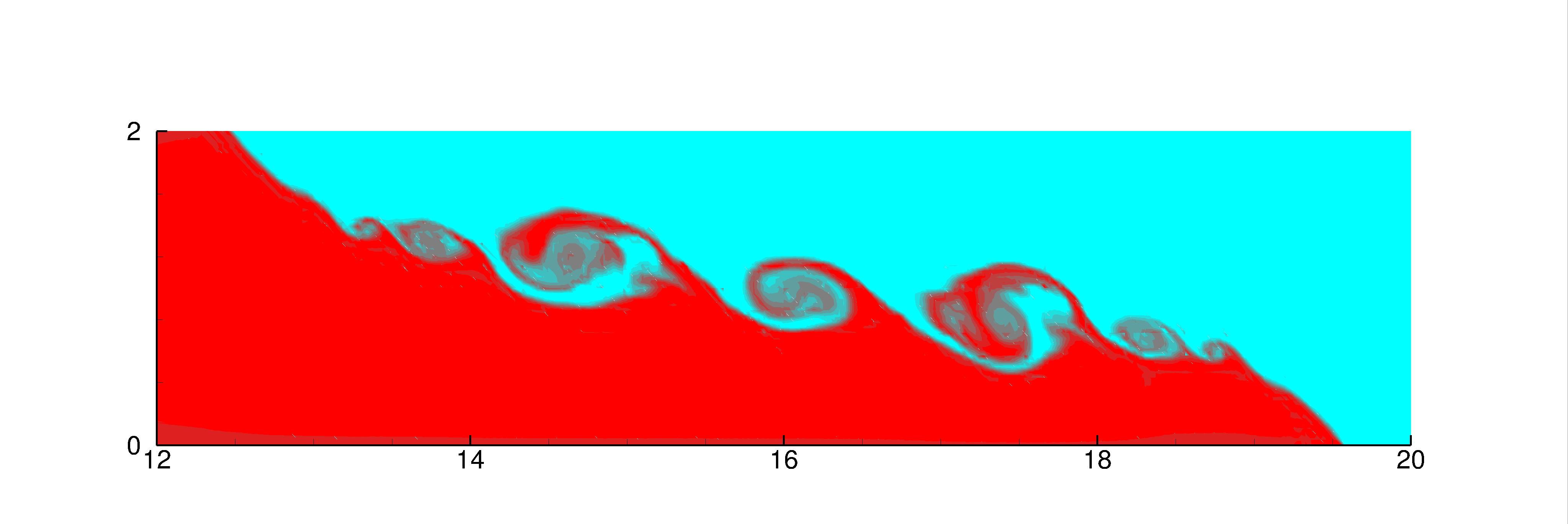}}
\end{subfigure} 

\begin{subfigure}[]{
      \label{fig:numflux_c}
\vspace{-14mm}
      \includegraphics[width=0.8\textwidth]{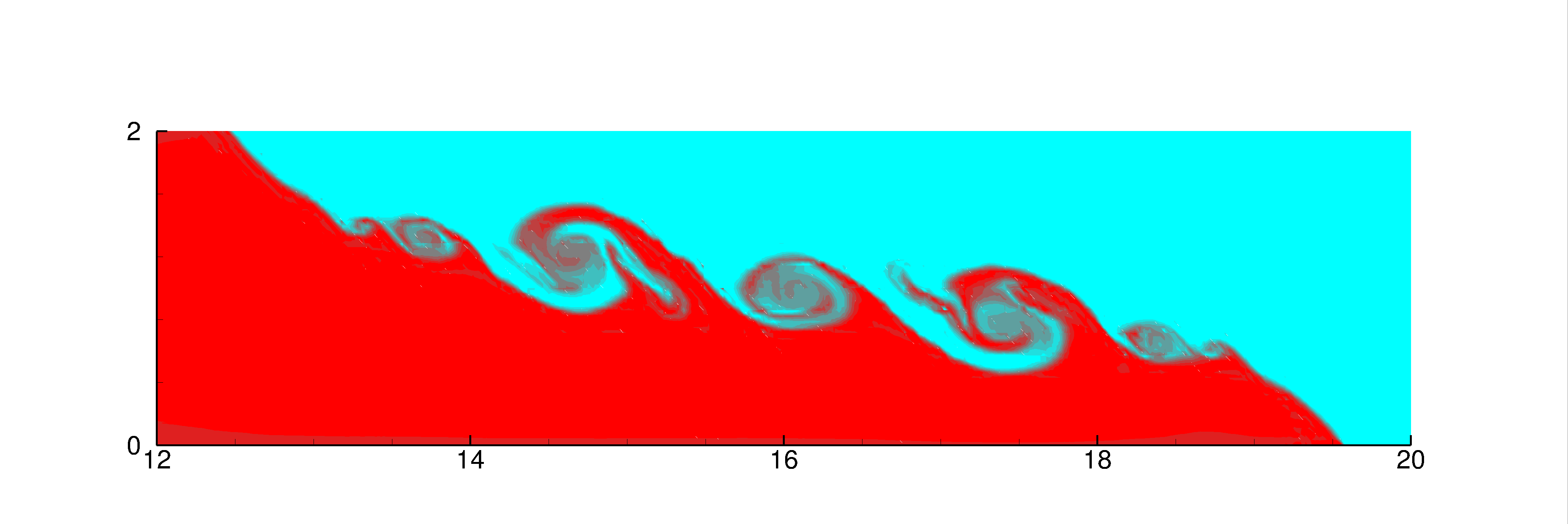}}
\end{subfigure}
\caption{Density for $Re=44721$ and $\gamma_r=0.96$ at time $t=5$. (a) Rusanov flux. (b) Godunov exact Riemann solver. (c) Modified Rusanov flux (upwinding velocity equal to the velocity of the fluid)}
\label{fig:numflux}
\end{figure}

In this section some results obtained with the Rusanov numerical flux, a modified version of the Rusanov flux (where the upwinding velocity is set equal to the velocity of the fluid) and the exact Godunov Riemann solver (implemented as in \cite{gottlieb:1988}) will be presented. In particular, in Figure \ref{fig:numflux} we have the density profiles of a Boussinesq simulation for $\RE=44721$ and $\gamma_r=0.96$, at $t=5$. We notice that the profile obtained with the Rusanov flux (Figure \ref{fig:numflux_a}) is highly inaccurate both for the incorrect reproduction of the shape and number of turbulent structures and for the presence of spurious oscillations. These spurious features are instead absent in the profiles obtained with the exact Godunov Riemann solver and with the modified version of the Rusanov flux (Figures \ref{fig:numflux_b} and \ref{fig:numflux_c}). This behaviour is probably due to the fact that, when dealing with low Mach number flows, the upwinding velocity in the Rusanov flux is equal to the sound velocity, which 
 is very different from the velocity of the fluid in this regime. As a consequence, upwinding is performed with a wrong velocity. This difficulty is indeed overcame by employing the exact solver or simply by substituting the upwinding velocity with the velocity of the fluid. This explanation appears to be consistent with the accurate results obtained by upwind-based, semi-Lagrangian schemes in the simulation of variable density flows with small density differences, see e.g. \cite{bonaventura:2000}, \cite{tumolo:2015}.

\section{Conclusions and future perspectives}
\label{conclu}
In this paper, we have assessed the capability of a DG-LES model for the numerical
simulation of turbulent gravity currents in the non-Boussinesq regime. 
  The results obtained by  application of the DG-LES  method 
  proposed in \cite{abba:2015}
  have been compared with those presented in \cite{birman:2005} in the incompressible case. 
The quality of our results with respect to \cite{birman:2005} has been first assessed using density contours at a fixed instant of time and for different density ratios $\gamma_r$. 
The density contours obtained with our DG simulations appears to be qualitatively similar to the ones in \cite{birman:2005} even though some  discrepancies are present, especially for the lower density ratio $\gamma_r=0.2$. 
 
From a more quantitative viewpoint, two different energy budgets have been considered. First of all, normalized values of potential, kinetic and dissipated energies as a function of time have been compared with the trends reported in \cite{birman:2005}, finding very good agreement.
The temporal evolution of the energy dissipated separately by the light and dense fronts  has  also been computed. Also in this case, we have quite good agreement with the results in \cite{birman:2005}, with a good prediction of the more dissipative nature of the dense front with respect to the light one. There is however a slight underestimation of the energy dissipated by the dense front. 
We can conclude that our approach is able to  reproduce quite well the incompressible results of \cite{birman:2005} in the   low Mach number regime. Some  preliminary LES results have also been presented. Even if these results have not been compared to corresponding DNS results, we can infer that the classic Smagorinsky model is probably too dissipative for the accurate simulation of turbulent gravity currents. On the other hand the results obtained with the dynamic model seem to indicate that backscatter plays a quite important role.
Concerning the impact of the numerical flux choice, we can conclude that the generally used Rusanov
flux may not be the best option  for variable density low Mach number regime. 

From the physical point of view, the next planned step is to carry out Large Eddy Simulations employing also more advanced models, like the anisotropic dynamic model proposed in \cite{abba:2001} and the novel proposals in \cite{germano:2014} for compressible, variable density flows. Moreover, we have planned also to compute additional diagnostics, like the available potential energy (\cite{winters:1995}, \cite{tseng:2001}), in order to better discriminate the performances of the different models, as   already done in \cite{ozgokmen:2009}, \cite{berselli:2011} and \cite{berselli:2014}.  From the  computational point of view, we are aware that employing an explicit time integration method in presence of low Mach numbers leads to computational inefficiency; the implementation of an implicit time integration method will be probably necessary in order to perform also 3D simulations. Furthermore,
a deeper investigation of the influence of the numerical flux on the solutions  in variable density low-Mach number would be required to better understand some of the results presented in this paper.

 \section*{Acknowledgements} 
 This paper contains an extended version of results presented by the first author at the 2016 SIMAI Congress
 and is part of the first author's PhD thesis work.
 We are happy to acknowledge the continuous help of M. Restelli and M.Tugnoli with the application of the FEMILARO code. Several useful discussions with T. Esposti Ongaro and M. Cerminara are also kindly acknowledged.
  The results of this research have been achieved using the 
computational resources made available   at CINECA (Italy) by the LISA high performance computing project
 {\it DECLES: Large Eddy Simulation of Density Currents and Variable Density Flows, HPL13PJ6YS}.
 
\bibliographystyle{plain}
\bibliography{lock_exchange}

\end{document}